# Reversal of Spin-torque Polarity with Inverting Current Vorticity in Composition-graded Layer at the Ti/W Interface

Hayato Nakayama[1], Taisuke Horaguchi[1], Jun Uzuhashi[2], Cong He[2], Hiroaki Sukegawa[2], Tadakatsu Ohkubo[2], Seiji Mitani[2], Kazuto Yamanoi[1], and Yukio Nozaki[1,3]*

[1]Department of Physics, Keio University, 3-14-1 Hiyoshi, Kohoku-ku, Yokohama, Kanagawa, 223-8522, Japan

[2]Research Center for Magnetic and Spintronic Materials, National Institute for Materials Science, 1-2-1 Sengen, Tsukuba, Ibaraki, 305-0047, Japan

[3]Center for Spintronics Research Network, Keio University, 3-14-1 Hiyoshi, Kohoku-ku, Yokohama, Kanagawa, 223-8522, Japan

E-mail*: nozaki@phys.keio.ac.jp

**Abstract**

While compositional gradient-induced spin-current generation has been explored, its microscopic mechanisms remain poorly understood. Here, the contribution of polarity of compositional gradient on spin-current generation is explored. A nanoscale compositional gradient, formed by in-situ atomic diffusion of ultrathin Ti and W layers, is introduced between 10-nm-thick W and Ti layers. Spin-torque ferromagnetic resonance in ferromagnetic $Ni_{95}Cu_5$ deposited on this gradient reveals that a moderate compositional gradient suppresses negative spin torque from the spin Hall effect in W. In contrast, reversing the Ti/W stacking order, which inverts the gradient, suppresses positive spin torque from the orbital Hall effect in Ti. These findings suggest that the sign of spin torque is governed by the polarity of compositional gradient, providing a novel strategy for efficient spin-torque generation without relying on materials with strong spin or orbital Hall effect.

## 1. Introduction

Spin current, the transfer of spin angular momentum by electrons, plays a crucial role in spintronics by enabling the manipulation of magnetization in ferromagnets through spin torque. This capability has led to the development of spintronic devices, such as magnetoresistive random access memory (MRAM). While Pt, $\beta$-W, or $\beta$-Ta thin films, known for their strong spin–orbit interaction (SOI), have been a conventional choice for generating spin torque, recent studies have shown that heterostructures with compositional gradients, such as surface-oxidized Cu films,[1,2] and Si/Al compositionally graded interfaces,[3] can achieve comparable spin-torque efficiencies using only light metals with weak SOI. These findings open the door to more sustainable spintronic technologies that avoid the use of rare metals. However, the underlying mechanisms driving spin-torque generation in these compositionally graded systems remain poorly understood.

Key mechanisms proposed to explain this phenomenon include the Rashba–Edelstein effect (REE),[4–7] orbital REE (OREE),[8, 9] and spin–vorticity coupling (SVC).[10–13] The REE and OREE, which arise from broken spatial inversion-symmetry, predict the strongest effects at sharp interfaces. In contrast, SVC, which does not rely on inversion-symmetry breaking, involves coupling between electron spin and macroscopic rotation via the gyromagnetic effect. Consequently, SVC-induced spin currents may not peak at the steepest compositional gradients. Evidence of spin-current generation via SVC has been observed in systems with lattice rotations induced by surface acoustic waves,[12, 14, 15] and nonuniform liquid metal flow,[13, 16] where vorticity coupled with electron spin arises from atomic motion. These phenomena have also been investigated theoretically.[10, 11, 17–19] Moreover, recent theoretical studies have shown that the current vortices interact with conduction electron spins via SVC, and that vorticity induces an effective magnetic field for localized spins through the *sd*-exchange interaction.[20] In particular, the effective magnetic field generated by vorticity arising from electric current flow around a notch structure can create skyrmions. Furthermore, both theoretical and experimental studies have demonstrated that the spatial gradient of vorticity drives skyrmions, generated in notch structures, via spin currents produced through SVC.[20,21] Notably, in these cases, the electric current flows within magnetic materials. Further experimental studies are necessary to clarify the role of this mechanism in spin-torque generation in compositionally graded systems.

In our previous study, we developed Ti/W/Ni-Cu gradient materials featuring a compositionally graded interface (CGI) formed by alternately stacking ultrathin Ti and W layers.

We quantified spin-torque efficiency using spin-torque ferromagnetic resonance (ST-FMR).[22] The CGI width was precisely controlled by adjusting the thickness of the W/Ti bilayer $t_i$ from 0 to 2 nm, with the width measured at 1.4 nm for $t_i = 0$ and 2.0 nm for $t_i = 1.0$ nm. Our measurements reveal that the spin-torque efficiency peaked at $t_i = 0.5$ nm, indicating a non-monotonic relationship between $t_i$ and spin-torque efficiency. Additionally, we observed pronounced nonreciprocal charge–spin conversion at the Ti/W CGI, supported by a nearly constant effective Gilbert damping constant across all samples. We hypothesize that SVC is responsible for the observed $t_i$ dependence and nonreciprocal behavior, drawing parallels with phenomena seen in surface-oxidized metals. As shown in **Figure 1**, the Ti/W/Ni-Cu gradient materials exhibit a variation in electrical conductivity along the film thickness, leading to a gradient in electron drift velocity. This gradient induces vorticity due to the nonuniform laminar flow of electrons. In SVC theory, the vorticity of velocity field, $\boldsymbol{\omega} = \nabla \times \boldsymbol{v}$, couples with the electron spin angular momentum $\boldsymbol{S}$ through the following Hamiltonian:[10]

$$H_{\text{SVC}} = -\frac{1}{2} \boldsymbol{S} \cdot \boldsymbol{\omega}. \tag{1}$$

The spin current arises due to the gradient of effective magnetic field, which is aligned with $\boldsymbol{\omega}$, exerting a spin-dependent force on the electrons. In compositionally graded materials, the vorticity is oriented in the in-plane direction and exhibits a gradient along the out-of-plane direction, generating a spin current with in-plane spin polarization flowing along the out-of-plane direction. This spin current flows into an adjacent ferromagnetic material, exerting torque on its magnetization via the conventional spin-transfer mechanism. Importantly, our approach separates the spin-current generation region from the detection region, allowing us to directly examine the coupling between vorticity and conduction electron spins. Understanding the microscopic mechanism underlying SVC-mediated spin-current generation driven by electric current vorticity is crucial, as it remains to be fully established theoretically.

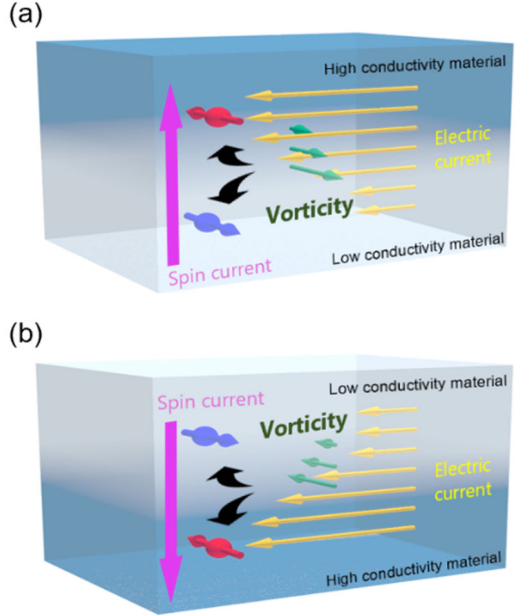

**Figure 1.** Charge-to-spin conversion in compositionally graded materials. Yellow arrows indicate the electric current, which varies along the film thickness due to the spatial distribution of electrical conductivity, resulting in electric current vorticity. Green arrows depict the effective magnetic field induced by this vorticity through gyromagnetic SVC. The spin current is generated along the gradient of the vorticity. a) In gradient materials such as Ti/W/Ni-Cu, where the bottom layer consists of a material with low electrical conductivity and the top layer a material with high electrical conductivity, an upward spin current is generated. b) In gradient materials such as W/Ti/Ni-Cu, where the bottom layer consists of a material with high electrical conductivity and the top layer a material with low electrical conductivity, a downward spin current is expected due to the opposite polarity of the vorticity compared to (a).

In this paper, we demonstrate the spin-torque modulation induced by electric current vorticity with a reversed polarity compared to previously reported Ti/W/Ni-Cu samples. To achieve this, we fabricated the CGI with Ti and W in a reverse stacking order resulting in W/Ti/Ni-Cu gradient materials, as shown in Figure 1(b). The CGI width was systematically controlled by varying the thickness of ultrathin Ti/W bilayer insertion. We evaluated spin-torque efficiency using ST-FMR measurements. This study aims to elucidate the role of the polarity of electric current vorticity in spin-current generation, offering insights that build upon and contrast with our earlier findings in Ti/W/Ni-Cu systems.

## 2. Results and Discussion

### 2.1. Structural Analyses of W/Ti CGIs

We fabricated microstrips with the structure of Ti(3 nm)/W(10 nm)/Ti($t_i$/2)/W($t_i$/2)/Ti(10 nm)/Ni$_{95}$Cu$_5$[Ni-Cu](10 nm), covered with a SiO$_2$(20 nm) cap, on thermally oxidized Si substrates using magnetron sputtering. The thickness $t_i$ of the ultrathin Ti/W bilayer was varied from 0 to 2.0 nm in intervals of 0.5 nm to facilitate atomic diffusion at the interface between W(10 nm) and Ti(10 nm) layers and form the CGI. The 3-nm-Ti layer acts as a seed layer for $\alpha$-W growth. **Figure 2** shows a schematic of the W/Ti/Ni-Cu gradient materials.

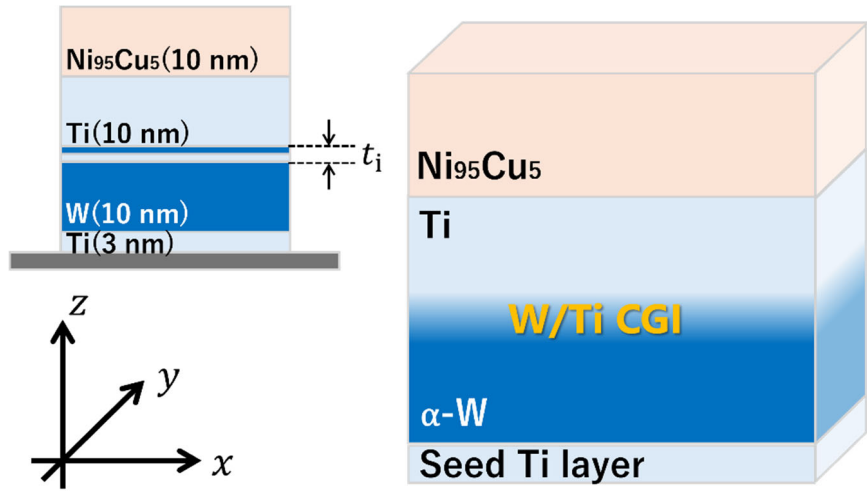

**Figure 2.** Schematic of the W/Ti/Ni-Cu gradient materials. Illustration of the designed (left) and actual (right) structures of the W/Ti/Ni-Cu gradient materials, along with the coordinate system used in this paper. The ultrathin Ti/W bilayer undergoes spontaneous mixing, forming a W/Ti CGI.

Two-dimensional X-ray diffraction (2D-XRD) measurements were conducted to analyze the crystal structure of the W/Ti/Ni-Cu gradient materials. Tungsten typically exhibits two phases in thin films: highly conductive $\alpha$-W with a body-centered cubic (BCC) structure and poorly conductive $\beta$-W with an A15 structure.[23–25] **Figure 3** shows XRD spectra for various $t_i$ samples. The W/Ti/Ni-Cu samples showed peaks corresponding to face-centered cubic (FCC) Ti, $\alpha$-W, and FCC Ni-Cu, confirming selective $\alpha$-W growth as seen in our earlier Ti/W/Ni-Cu systems.[22]

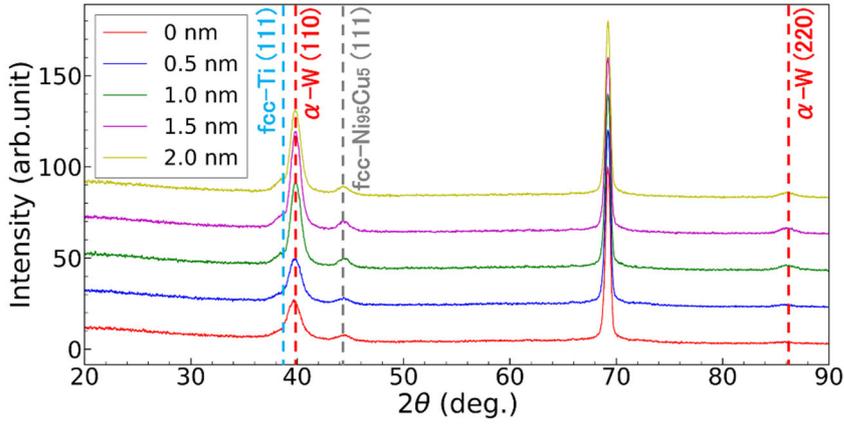

**Figure 3.** 2D-XRD spectra for W/Ti/Ni-Cu gradient materials with $t_i$ = 0 nm (red), 0.5 nm (blue), 1.0 nm (green), 1.5 nm (magenta), and 2.0 nm (yellow). All measurements were performed using Cu $K\alpha$ radiation (wavelength = 1.54 Å). Note that a diffraction peak appearing at around $2\theta$ = 69 deg. is caused by the thermally oxidized Si substrate.

Surface roughness was assessed using atomic force microscopy (AFM). **Figure 4**a shows a 25 × 25 μm² AFM image of the $t_i$ = 1.0 nm sample. Island structures with diameters from 0.5 μm to 3 μm and heights around 1.5 nm were observed. No such structures were present for $t_i$ = 0 and 0.5 nm samples. We have already verified such island structures prevent spin-current generation at the CGI.[22] Compared to our earlier Ti/W/Ni-Cu gradient materials, the W/Ti/Ni-Cu sample had fewer island structure. In addition, Figure 4b, showing a higher magnification AFM image (5 × 5 μm²), confirms a relatively smooth surface, hence spin-current generation at the CGI will not be disturbed. The formation of island-like structure is attributed to the rapid sputtering rate (see Section 4, Experimental Section). Root-mean-square roughness (RMSR) and peak-to-valley ($R_{pv}$), which represent the standard deviation of the height distribution and the difference between the highest peak and the lowest valley, respectively, were evaluated from AFM images (Figure 4b). Figure 4c,d illustrates the RMSR and $R_{pv}$, respectively, as a function of $t_i$. The RMSR and $R_{pv}$ were comparable to those of Ti/W/Ni-Cu samples where CGI-mediated spin-torque generation was observed.

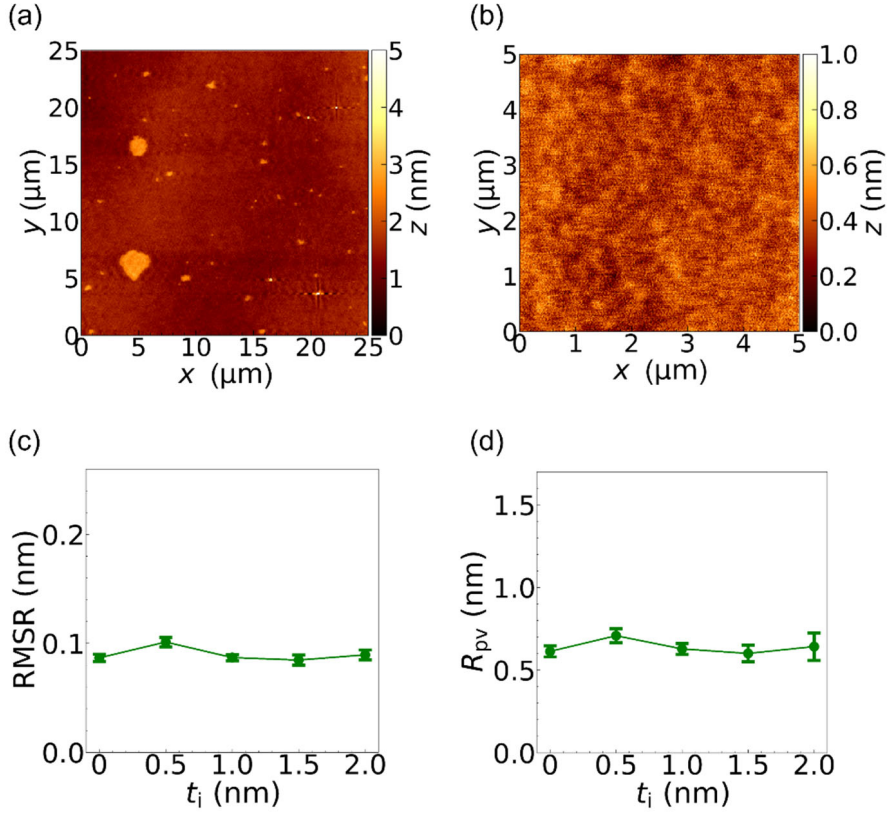

**Figure 4.** AFM images and quantitative roughness analysis for W/Ti/Ni-Cu gradient materials. a,b) AFM images of the W/Ti/Ni-Cu samples with $t_i = 1.0$ nm, showing scan size of (a) 25 × 25 µm² and (b) 5 × 5 µm². c,d) (c) RMSR and (d) $R_{pv}$ for W/Ti/Ni-Cu gradient materials. RMSR and $R_{pv}$ values are averaged over the 16 squares of 1.25 × 1.25 µm², derived from the AFM scans.

The cross-sectional bright field scanning transmission electron microscopy (BF-STEM) with energy-dispersive X-ray spectroscopy (EDS) was performed for samples with $t_i = 0$, 1.0, and 1.5 nm to analyze the CGI formation. **Figure 5** shows the BF-STEM images, nanobeam diffraction patterns, and the corresponding EDS line profiles. As expected from the AFM images (Figure 4), no structural undulations were observed. Continuous brightness variation in Figures 5a,d ($t_i = 0$ nm) and 5b,e ($t_i = 1.0$ nm) indicates simple gradients at the W(10 nm)/Ti(10 nm) interface. The CGI widths measured from 75% compositional ratios were 0.97 and 1.94 nm for the samples with $t_i = 0$ (Figure 5g) and 1.0 nm (Figure 5h), respectively. In contrast, Figures 5c and 5f ($t_i = 1.5$ nm) reveal alternating bright and dark regions, suggesting an alloyed $Ti_{40}W_{60}$ layer approximately 1 nm thick. Despite this intermixing, two separate CGIs —W/$Ti_{40}W_{60}$ and $Ti_{40}W_{60}$/Ti—are present, though with an effective CGI thickness of less than 1.88 nm. Moreover, our previous study reveals that the electrical resistivity of Ti–W system changes almost linearly with respect to W concentration,[22] indicating that the polarity of electric current vorticity remains consistent across all samples. Thus, while the precise nature of the interface requires further investigation, the role of CGIs in spin-torque generation can still be reliably discussed. We also confirmed the existence of a 1.2 nm compositional gradient at the Ti(3 nm)/W(10 nm) interface. The spin current it may generate is discussed in Section 2.3, where we show that the bulk effect of Ti(3 nm) and W(10 nm), together with this CGI.

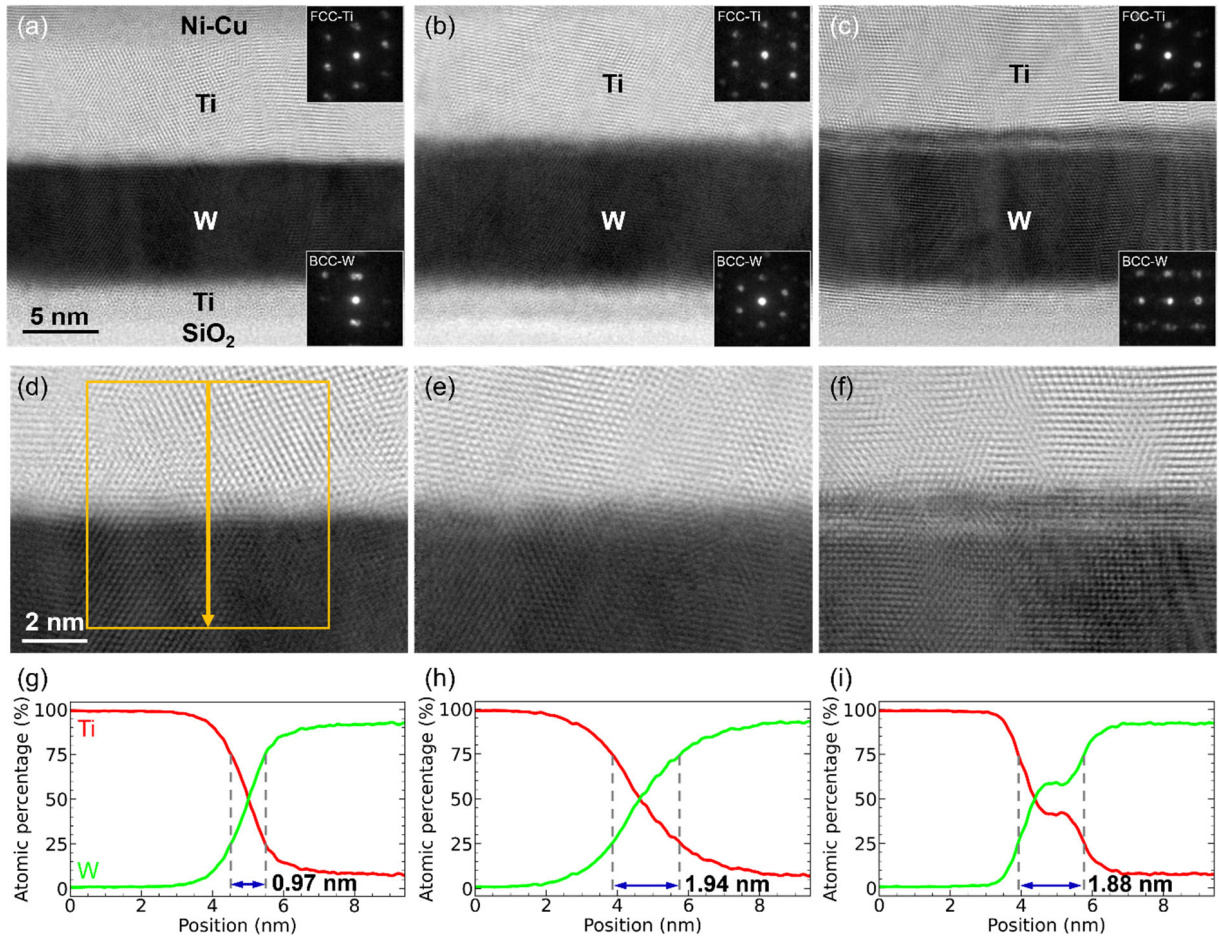

**Figure 5.** BF-STEM images and EDS profiles of W/Ti/Ni-Cu gradient materials: $t_i = 0$ (a,d,g), $t_i = 1.0$ nm (b,e,h), and $t_i = 1.5$ nm (c,f,i). The orange dashed box in (d) highlights the EDS integration area.

## 2.2. Spin-torque Efficiency Measurement and Analysis

We conducted ST-FMR measurement to evaluate the spin-current generation in the W/Ti/Ni-Cu gradient materials.[26–31] The experimental setup is illustrated in **Figure 6**a. A signal generator (SG) was connected to the multilayer microstrip to apply alternating current (AC), and a nanovoltmeter (NVM) was used to detect the direct current (DC) voltage generated in the microstrip through a broadband bias tee. When an AC current $\boldsymbol{j}_{\text{rf}}$ is applied to the microstrip, it generated both an alternating Oersted field $\boldsymbol{h}_{\text{Oe}}$ and a spin current $\boldsymbol{j}_{\text{s}}$, which exert field torque and spin torque, respectively, on the magnetization $\boldsymbol{m}$ of the ferromagnetic Ni-Cu layer. These torques induce precession of $\boldsymbol{m}$. The rectified DC voltage $V_{\text{DC}}$, which varies due to the anisotropic magnetoresistive effect, was measured while sweeping the in-plane external magnetic field $\boldsymbol{B}$ at a frequency and power of $\boldsymbol{j}_{\text{rf}}$ up to 20 GHz and 20 dBm, respectively. The field $\boldsymbol{B}$ was applied in the $x$-$y$ plane at a 45-degree angle relative to the $x$ axis.

The spin-torque efficiency $\xi_{\text{FMR}}$ was determined from the rectified DC voltage $V_{\text{DC}}$. The $V_{\text{DC}}$ spectrum as a function of $B$ was decomposed into Lorentzian $F_{\text{L}}(B)$ and anti-Lorentzian $F_{\text{AL}}(B)$ components as follows:

$$V_{\text{DC}} = V_{\text{L}} F_{\text{L}}(B) + V_{\text{AL}} F_{\text{AL}}(B) + V_0, \qquad (2)$$

where $F_{\text{L}}(B) = \Delta^2/\{(B - B_{\text{r}})^2 + \Delta^2\}$ and $F_{\text{AL}}(B) = \Delta(B - B_{\text{r}})/\{(B - B_{\text{r}})^2 + \Delta^2\}$. Here, $\Delta$ is the linewidth of the ST-FMR spectrum, $B_{\text{r}}$ is the resonant field, and $V_0$ is the offset voltage. $\xi_{\text{FMR}}$ was calculated using the amplitude ratio between $V_{\text{L}}$ and $V_{\text{AL}}$:[26, 30]

$$\xi_{\text{FMR}} = \frac{V_{\text{L}}}{V_{\text{AL}}} \frac{e \mu_0 M_s d_{\text{FM}} d_{\text{NM}}}{\hbar} \left(1 + \frac{\mu_0 M_s}{B_{\text{r}}}\right)^{\frac{1}{2}}, \qquad (3)$$

where $d_{\text{FM}}$ and $d_{\text{NM}}$ are the thicknesses of the ferromagnetic and nonmagnetic layers, respectively. Here, $\hbar$ is the reduced Planck constant, $e$ is the elementary charge, and $\mu_0$ is the magnetic constant. The saturation magnetization $\mu_0 M_s$ was derived from the Kittel formula:

$$f = \frac{\gamma}{2\pi} \sqrt{B_{\text{r}}(B_{\text{r}} + \mu_0 M_s)}, \qquad (4)$$

where $f$ and $\gamma$ are the resonant frequency and the gyromagnetic ratio, respectively.

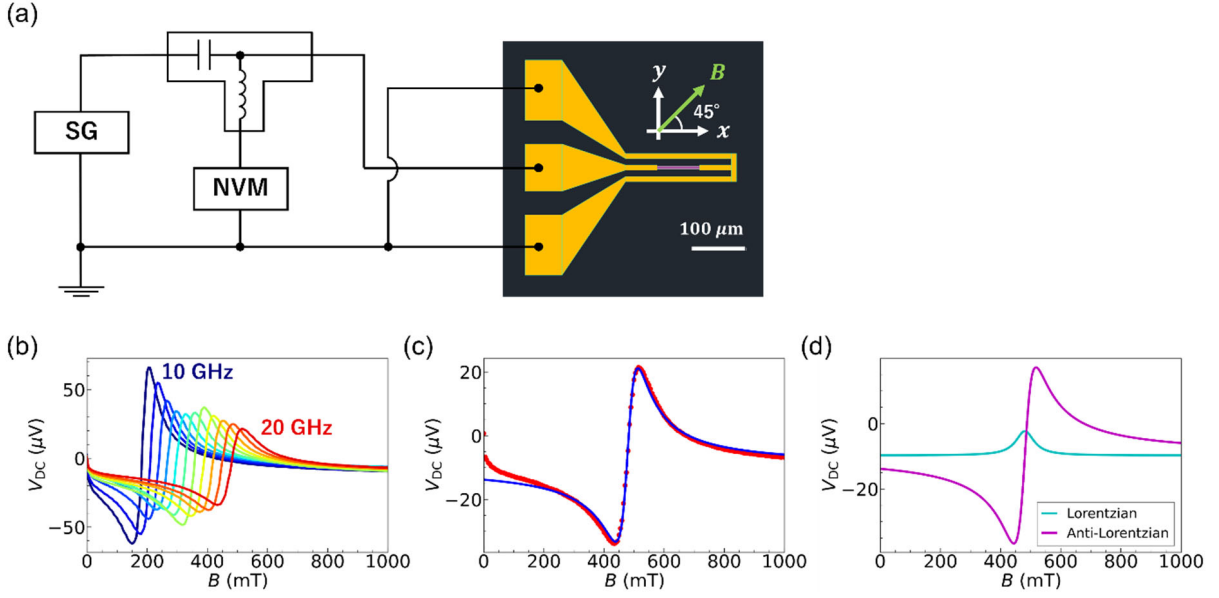

**Figure 6.** Spin-torque ferromagnetic resonance measurement setup and results. a) Schematic of experimental setup for ST-FMR measurement. The microstrip dimensions were $100 \times 5$ μm$^2$, and the external magnetic field ***B*** was applied in the $x$-$y$ plane at a 45-degree-angle relative to the $x$ axis. b) Raw ST-FMR spectra for a W/Ti/Ni-Cu gradient material with $t_i$ = 1.0 nm, at frequencies ranging from 10 GHz (solid navy curve) to 20 GHz (solid red curve). c) ST-FMR spectrum at 20 GHz (solid red circles), fitted using Equation 2 (solid blue curve). d) Decomposed Lorentzian (solid cyan curve) and anti-Lorentzian (solid magenta curve) components of the fit.

## 2.3. Evaluation of Spin-torque Efficiency

Figure 6b shows the ST-FMR spectrum for the W/Ti/Ni-Cu sample with $t_i$ = 1.0 nm as a representative example. The blue solid curve in Figure 6c shows the best fit using Equation 2, allowing us to separate the Lorentzian and anti-Lorentzian components, $F_L(B)$ and $F_{AL}(B)$, respectively. These components are individually illustrated in Figure 6d. Using the amplitudes $V_L$ and $V_{AL}$ of $F_L(B)$ and $F_{AL}(B)$, we evaluated $\xi_{FMR}$ for all W/Ti/Ni-Cu samples according to Equation 3, with the results summarized in **Figure 7**a. Each $\xi_{FMR}$ value for W/Ti/Ni-Cu samples in Figure 7a represents the average result from up to six samples prepared on the same substrate.

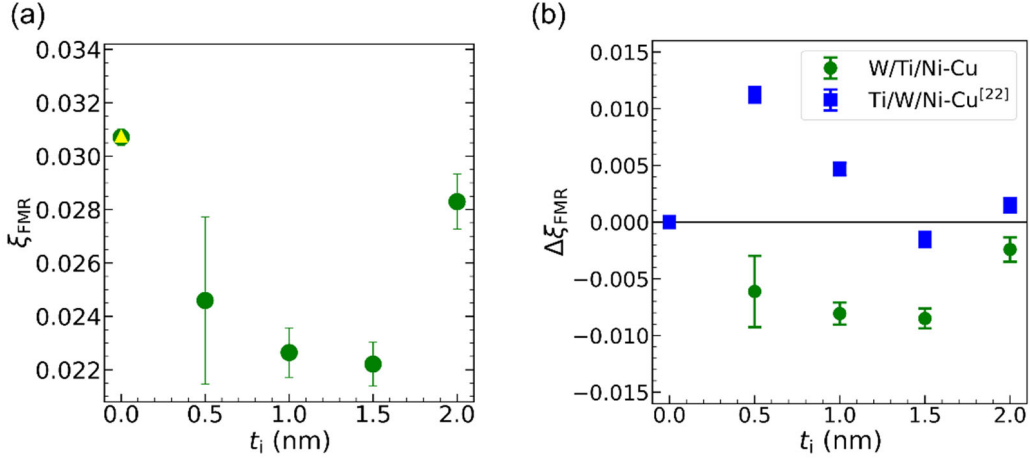

**Figure 7.** a) Dependence of spin-torque efficiency on insertion layer thickness. Spin-torque efficiency $\xi_{FMR}$ as a function of insertion layer thickness $t_i$, derived from ST-FMR measurement at 20 GHz (green circle plots). Each error bar represents standard deviation of $\xi_{FMR}$ for up to six samples on the same substrates. Yellow triangle plot represents $\xi_{FMR}$ of the reference Ti(10 nm)/Ni-Cu(10 nm) sample.[22] b) Comparison of spin-torque efficiency between Ti/W/Ni-Cu and W/Ti/Ni-Cu gradient materials. Difference of the change in spin-torque efficiency $\Delta\xi_{FMR}$ (defined as $\xi_{FMR}(t_i) - \xi_{FMR}(t_i = 0 \text{ nm})$) as a function of insertion layer thickness $t_i$ for Ti/W/Ni-Cu gradient materials[22] (solid blue squares) and W/Ti/Ni-Cu gradient materials (solid green circles, this work).

First, we investigate the bulk effects of W and Ti, focusing on $t_i$ = 0 nm sample, which consists of a Ti(3 nm)/W(10 nm)/Ti(10 nm)/Ni-Cu(10 nm) multilayer. In our previous study on Ti/W/Ni-Cu gradient materials,[22] we observed a negative $\xi_{FMR}$ for the $t_i$ = 0 nm sample. We attribute this result to the interplay between negative spin current generated by the bulk spin Hall effect (SHE) in W and a positive orbital current produced by the bulk orbital Hall effect (OHE) in Ti. Thus, it is expected that the SHE in W and the OHE in Ti similarly influence the ferromagnetic Ni-Cu layer in the W/Ti/Ni-Cu gradient materials. As shown in Figure 7a, $\xi_{FMR}$ for the $t_i$ = 0 nm sample is +0.031, a value comparable to that of the previously reported Ti(10 nm)/Ni-Cu(10 nm) bilayer,[22] depicted by an yellow triangle plot. This suggests that the OHE in Ti(10 nm) is the dominant contributor to this result. In other words, the bulk effect of the Ti(3 nm) and W(10 nm) layers, together with this CGI largely cancels out, making its net contribution negligible for the measured $\xi_{FMR}$. Moreover, since this interface remains unchanged for all samples, it does not affect the $t_i$-dependent CGI effects at the W(10 nm)/Ti(10 nm) interface, which dominate the $\xi_{FMR}$ variation.

Next, we analyze the dependence of $\xi_{FMR}$ on $t_i$. As $t_i$ increases from 0 to 0.5 nm, $\xi_{FMR}$ decreases from 0.031 to 0.025, and continues to decrease as $t_i$ increases further followed by rapid increase in $\xi_{FMR}$ at $t_i = 2.0$ nm, i.e., the CGI effect increased up to $t_i = 1.5$ nm, and it weakened at $t_i = 2.0$ nm. This trend indicates that the insertion of the W/Ti CGI generates negative spin torque. Notably, the SHE and OHE in W and Ti, respectively, remain independent of $t_i$, as confirmed by the consistent XRD spectra with varying $t_i$ (Figure 3). Additionally, the suppression of orbital current transport, which is sensitive to crystallographic defects,[32] remains consistent across all samples since the interfacial roughness between Ti(10 nm) and Ni-Cu(10 nm) does not show significant variation, as validated by AFM analysis (see Figure 4c,d). Therefore, the decrease in $\xi_{FMR}$ can be directly linked to spin-current generation at the CGI. Furthermore, the standard deviation of $\xi_{FMR}$ for $t_i = 0.5$ nm is notably large. Our previous findings indicate that $\xi_{FMR}$ modulation was more pronounced in Ti/W/Ni-Cu gradient materials with smaller $t_i$. From this perspective, the increased deviation of $\xi_{FMR}$ for smaller $t_i$ may be attributed to slight differences in CGI width within the same substrate, leading to significant variations in $\xi_{FMR}$.

### 2.4. Reversal of Spin-torque Polarity with Inversion of the Ti/W CGI

Here, we discuss gyromagnetic SVC as the mechanism responsible for spin-torque generation at the CGI. SVC-mediated spin-torque generation has been previously reported in surface-oxidized Cu,[2] and Ti/W/Ni-Cu gradient materials.[22] Comparing the $t_i$-dependence of $\xi_{FMR}$ in W/Ti/Ni-Cu gradient materials with that in Ti/W/Ni-Cu materials offers insights into SVC-mediated spin-current generation.[22] As shown in Figure 7b, the insertion of an ultrathin Ti/W or W/Ti bilayer resulted in a decrease in $\xi_{FMR}$ in W/Ti/Ni-Cu gradient materials (green circles), whereas it led to an increase in Ti/W/Ni-Cu materials (blue squares). The difference between W/Ti/Ni-Cu and Ti/W/Ni-Cu gradient materials in the polarity of the electric current vorticity is due to the relative conductivities of Ti (low) and W(high), adjacent to the ferromagnetic Ni-Cu. These findings suggest that the sign of SVC-mediated spin current is determined by the polarity of electric current vorticity.

However, as shown in Figure 7b, the $t_i$ value, at which the absolute value of $\Delta\xi_{FMR}$ reaches its maximum, varies depending on the stacking order of the Ti and W layers. In the previous ST-FMR measurement on Ti/W/Ni-Cu gradient materials, the insertion of a W/Ti layer increased $\xi_{FMR}$. Blue squares in Figure 7b reveal a significant positive increase in $\xi_{FMR}$ within narrower CGI, although this increase disappears at $t_i$ = 1.5 nm, where the ultrathin W/Ti bilayer transitions from a CGI to a distinct, separated structure. Contrastingly, the reversely stacked W/Ti/Ni-Cu gradient materials exhibited a different trend, with $\xi_{FMR}$ peaking not at $t_i$ = 0.5 nm but at $t_i$ = 1.5 nm. As shown in Figure 5i, the $Ti_{40}W_{60}$ layer is about 1 nm thick in the $t_i$ = 1.5 nm sample, implying that two separate CGIs —W/$Ti_{40}W_{60}$ and $Ti_{40}W_{60}$/Ti—are present. Because the Ti–W system exhibits a monotonous variation in resistivity with respect to W composition,[22] both interfaces likely contribute to negative spin-current generation via SVC, akin to the $t_i$ = 1.0 nm sample. Their combined effect yields a minimum value in $\xi_{FMR}$ at $t_i$ = 1.5 nm.

Furthermore, we examined the sputtering condition dependence of spin-torque efficiency as a function of $t_i$ for W/Ti/Ni-Cu samples. We reduced the sputtering rate for the Ti layer from 0.039 ± 0.001 nm/s to 0.022 ± 0.001 nm/s, matching the rate used in the growth of previous Ti/W/Ni-Cu samples that exhibit a CGI effect,[22] while maintaining the sputtering rate for the W layer at 0.046 ± 0.001 nm/s. Under these conditions, spin-torque efficiency remains nearly constant (+0.031 ~ +0.034) across $t_i$, suggesting that a pronounced CGI was not formed. This distinction arises from the difference in the CGI properties between the previous case—where the heavy metal W was deposited onto the light element Ti—and the current case—where the light element Ti is deposited onto the heavy element W. The deposition conditions required to achieve the CGI effect should be optimized according to the stacking order of nonmagnetic layers that compose the gradient material.

In addition to the SVC effect, we also need to consider the bulk spin Hall effect of the Ti–W nanoalloy formed within the CGI. The ST-FMR results suggest that the bulk effects in the Ti–W nanoalloy could contribute to the generation of negative spin torque, as evidenced by the decreasing $\xi_{FMR}$ with increasing $t_i$, which correlates with the expanding volume of the Ti–W nanoalloy. Indeed, about 1 nm of $Ti_{40}W_{60}$ alloy was observed in the $t_i$ = 1.5 nm sample. However, our data shows that $\xi_{FMR}$ for the Ti–W nanoalloy, evaluated using Sub./[Ti(0.5 nm)/W(0.5 nm)]$_{10}$/Ni-Cu(10 nm) structure, is positive at 0.026. These results indicate that negative spin-current generation due to the bulk effects at the CGI is unlikely.

Finally, we discuss the impact of REE and OREE on spin-torque efficiency. In this context, we must consider two interfaces that can contribute to spin-torque generation: Ti(10 nm)/Ni-Cu(10 nm) and W(10 nm)/Ti(10 nm). In the former case, REE and OREE predominantly produce field-like torque;[33] however, the magnitude of this torque is independent of $t_i$ since this interface remains consistent across all samples. When discussing the $t_i$-dependence of spin-torque efficiency, REE and OREE at the W(10 nm)/Ti(10 nm) interface should be examined. As indicated by the yellow triangle plot in Figure 7a, the $\xi_{FMR}$ value for the $t_i = 0$ nm sample (+0.031) is comparable to that of the Ti(10 nm)/Ni-Cu(10 nm) bilayer. This suggests that the sharpest W/Ti interface does not contribute to spin-torque generation. Therefore, we conclude that REE and OREE, both of which are caused by inversion symmetry breaking, are not key factors in this system. These points reinforce our conclusion that SVC is the most likely source of spin-current generation in these materials.

## 2.5. SVC-Mediated Spin Torque Magnitude

Finally, we discuss the magnitude of the SVC-mediated spin torque. As shown in Figure 7b, the maximum variation in $\Delta\xi_{FMR}$ was 0.011 for the Ti/W CGI and 0.008 for the W/Ti CGI in this study. These values are approximately one-third of the orbital Hall torque of Ti ($\xi_{FMR} = 0.031$ for a Ti(10 nm)/Ni-Cu(10 nm) bilayer) and about one-seventh of the spin Hall torque of Pt ($\xi_{FMR} = 0.066$ for a Pt(10 nm)/Ni-Cu(10 nm) bilayer), indicating that the CGI effect in the Ti–W system is relatively small. We attribute this to moderate conductivity contrast between Ti and W, which is approximately threefold. Since electric current vorticity is generated by the spatial variation in electrical conductivity along the film thickness, the conductivity gradient plays a critical role. By comparison, a CGI comprising semiconductor Si and metal Al, which exhibit a much larger conductivity difference, achieved a damping-like torque efficiency of 0.67, exceeding that of Pt (0.21), despite both Si and Al being light elements.[3] This finding highlights that the SVC-mediated torque magnitude strongly depends on the material pair used to create the CGI. Importantly, materials with minimal bulk SHE of OHE can still enable efficient spin current generation if the CGI is designed to feature a significant conductivity gradient.

## 3. Conclusion

In this work, we successfully fabricated W/Ti/Ni-Cu gradient materials with varying electrical conductivity gradient along the film thickness and evaluated spin-torque efficiency via ST-FMR in the adjacent ferromagnetic Ni$_{95}$Cu$_5$ layer. Our investigation yielded three pivotal insights: First, the insertion of an ultrathin Ti/W bilayer with thickness $t_i$ does not result in the formation of a bulk Ti-W nanoalloy but instead modifies the interfacial structure between the W(10 nm) and Ti(10 nm) layers for $t_i \leq 1.0$ nm. When a Ti–W nanoalloy is formed for thicker $t_i$ samples, the difference in the sign of the spin-torque efficiency indicates that the bulk effect of the Ti–W nanoalloy does not play a primary role. Second, the SVC, which facilitates interaction between electron spin and macroscopic angular momentum, plays a critical role in spin-current generation in systems exhibiting compositional gradients. Notably, the electric current vorticity emerging at the CGI due to the conductivity gradient interacts with the electron spin within the W/Ti/Ni-Cu gradient materials. Third, the polarity of the spin current generated through SVC is dictated by the direction of the electric current vorticity. This understanding of SVC-driven spin-torque generation in materials with compositional gradients opens new possibilities for sustainable spintronic devices, expanding material options as electric current vorticity can be induced in any system with a conductivity gradient.

## 4. Experimental Section

*Preparation of W/Ti/Ni-Cu Gradient Materials*: We fabricated $100 \times 5$ μm$^2$ microstrips composed of Ti(3 nm)/W(10 nm)/Ti($t_i$/2)/W($t_i$/2)/Ti(10 nm)/Ni$_{95}$Cu$_5$(10 nm)/SiO$_2$(20 nm) on thermally oxidized Si substrates using a standard liftoff technique combined with magnetron sputtering. The CGI was engineered by leveraging the atomic intermixing that naturally occurs when sputtered particles, retaining high kinetic energy, impact the predeposited layers. By varying the thickness of the insertion layer $t_i$ from 0 to 2.0 nm in intervals of 0.5 nm, we controlled the CGI width between the W(10 nm) and Ti(10 nm) layers. The 3-nm-thick Ti layer serves as a seed layer to promote the growth of highly conductive α-W, facilitating a direct comparison of spin-torque generation in the W/Ti/Ni-Cu gradient materials with the previously studied Ti/α-W/Ni-Cu gradient materials.[22] The sputtering rates for Ti, W, and Ni$_{95}$Cu$_5$ were $0.039 \pm 0.001$ nm/s, $0.046 \pm 0.001$ nm/s, and $0.045 \pm 0.001$ nm/s respectively. To enhance the mixing of Ti and W at the interface, particularly given the mass disparity between the lighter Ti atoms and the heavier W atoms, the sputtering rate of Ti was increased relative to that used in prior studies of Ti/α-W/Ni-Cu samples.

*STEM observations*: The cross-sectional bright field (BF-) scanning transmission electron microscopy (STEM) observation with energy dispersive X-ray spectroscopy (EDS) and nanobeam electron diffraction was carried out at 300 kV accelerating voltage using Spectra Ultra S/TEM (Thermo Fisher Scientific). To guarantee the quality of data comparisons, the same 25-nm-thickness TEM lamellae were prepared by a thickness-controllable program using focused-ion-beam (FIB)-SEM dual-beam Helios5UX (Thermo Fisher Scientific) while considering FIB-damage becomes minimized.[34,35]

**Acknowledgements**

The authors thank T. Funato, M. Matsuo, and J. Fujimoto for theoretical discussions. A part of this work was supported by the Electron Microscopy Unit, National Institute for Materials Science (NIMS). This work was partially supported by the JST-CREST Program (no. JPMJCR19J4), the Grant-in-Aid for JSPS Fellows (no. 24KJ1955), the Grants-in-Aid for Scientific Research (no. 21H04565, 24H00322, 24H02233), the Grant-in-Aid for Research Activity Start-up (no. 22K20359), and the Spintronics Research Network of Japan (Spin RNJ).

**Conflict of Interest**

The authors declare no conflict of interest.

**Author Contributions**

H.N., T.H., and Y.N. planned to the study and wrote the manuscript. H.N. performed film deposition and fabricated the devices. J.U. and C.H. performed the STEM experiments with the help of H.S., S.M., and T.O. H.N carried out the measurements and analyzed the data with the help of T.H., K.Y., and Y.N. All the authors discussed the results and commented on the manuscript.